\documentclass[letter]{aa} 
\usepackage{graphicx,hyperref}
\usepackage{txfonts}
\usepackage{epstopdf}
\usepackage{longtable}
\usepackage{lscape}
\usepackage{natbib}
\usepackage{color}
\newcommand{\scs}{\scriptsize}
\begin{document} 
  \title{HD~16771: A lithium-rich giant in the red-clump stage}
   \author{Arumalla B. S. Reddy
          \inst{1}
          \and
          David L. Lambert\inst{1}
          }
   \institute{$^{1}$\,W.J. McDonald Observatory and Department of Astronomy, The University of Texas at Austin, Austin, TX 78712-1205, USA \\
              \email{bala@astro.as.utexas.edu}
             }
\date{}
  \abstract
{}   
{We report the discovery of a young lithium rich giant, HD 16771, in the core-helium burning phase that does not seem to fit existing proposals of Li synthesis near the luminosity function bump or during He-core flash. We aim to understand the nature of Li enrichment in the atmosphere of HD~16771 by exploring various Li enhancement scenarios.} 
{We have collected high-resolution echelle spectra of HD 16771 and derived stellar parameters and chemical abundances for 27 elements by either line equivalent widths or synthetic spectrum analyses.}   
{HD 16771 is a Li-rich (log\,$\epsilon$(Li)=$+2.67\pm0.10$ dex) intermediate mass giant star ($M=2.4\pm0.1\,M_{\odot}$) with age\,$\sim0.76\pm0.13$ Gyr and located at the red giant clump. Kinematics and chemical compositions are consistent with HD 16771 being a member of the Galactic thin disk population. The non-detection of $^{6}$Li ($<$\,3\%), a low carbon isotopic ratio ($^{12}$C/$^{13}$C=12$\pm$2), and the slow rotation ($v$\,sin$i$=2.8 km s$^{-1}$) all suggest that lithium might have been synthesized in this star. On the contrary, HD~16771 with a mass of 2.4$~M_{\odot}$ has no chance of encountering luminosity function bump and He-core flash where the possibility of fast deep-mixing for Li enrichment in K giants has been suggested previously.}
{Based of the evolutionary status of this star, we discuss the possibility that $^{7}$Li synthesis in HD~16771 is triggered by the engulfment of close-in planet(s) during the RGB phase.}


\authorrunning{A. B. S. Reddy and D. L. Lambert}
\titlerunning{Li-rich clump giant HD~16771}

\maketitle
%

\section{Introduction} 
Lithium depletion in giants is a natural consequence of stellar evolution. Soon after stars leave the main sequence the outer layers expand and cool, a star's rotation rate slows, the convection zone deepens and sweeps to the surface material from internal layers. This first dredge-up causes an overall dilution of Li abundance by about 1-2 dex (\citealt{Iben1967a},b). Indeed, abundance analysis of G and K giants \citep{Brown1989} even show much lower Li abundances in many giants than the expectations of standard stellar models (\citealt{Iben1967a},b) indicative of significant Li depletion during the pre-main and/or main sequence phases, and to some extent due to non-canonical mixing on the red giant branch (RGB). At the end of first dredge-up, significant abundance alterations to the giant's atmosphere are the depletion of fragile Li, increase in $^{4}$He, $^{14}$N, $^{13}$C, and dilution of $^{12}$C thereby lowering the $^{12}$C/$^{13}$C ratio down to $\sim$ 20$-$30 from its initial value of 90 on the main sequence \citep{Charbonnel1998}.

In spite of this convective mixing, a fraction of giants in the Galactic disk exhibit abnormally high Li abundances exceeding by nearly 2-3 orders of magnitude the standard model predictions for stars on the RGB\footnote{K giants with log\,$\epsilon$(Li)$>1.5$ dex are treated as Li-rich while those with Li abundances exceeding the present interstellar medium abundance, log\,$\epsilon$(Li)$>3.1$ dex, are termed as super Li-rich.} 
\citep{Brown1989,CharBal2000,Kumar2011}, and a few of them present rapid rotation (see \citealt{Carlberg2012} and references therein), strong infrared (IR) excesses \citep{delaReza1996}, and anomalously low $^{12}$C/$^{13}$C ratios \citep{Sneden1986}. Proposals that giants synthesize $^{7}$Li via Cameron-Fowler \citep{CamFow1971} $^{7}$Be-transport mechanism near RGB bump in low mass stars \citep{CharBal2000} and/or during the He-core flash in stars of M$<$2.2\,M$_{\odot}$ \citep{Kumar2011} have been challenged by the discovery of super Li-rich giants at magnitudes that span the RGB \citep{Monaco2011,MartellShetrone2013}. These observations imply that Li enrichment may not be limited to specific stellar evolutionary stages. Numerous models proposed hitherto are far from reaching a consensus on the synthesis of Li in K giants which may partly be attributed to the paucity of data on Li-rich K giants (see e.g. \citealt{Monaco2011} and \citealt{Carlberg2012} and references therein). 

Here we report the detection of a warm Li-rich clump-giant, a few of them found previously by \citet{Brown1989} and \citet{MartellShetrone2013}, whose mass and location in the Hertzsprung-Russell (HR) diagram is incompatible with an encounter with the luminosity function bump (LFB) or He-core flash where the internal production of Li in giants was proposed previously \citep{CharBal2000,Kumar2011}. Here, we restrict ourselves to discussing the chemical composition of HD 16771 and its evolutionary state in the HR diagram with an emphasis on providing a plausible scenario for the observed Li enrichment.

\section{Observations and data reduction}
The sample red giant star HD 16771 analysed in this letter was selected as a part of the program dedicated to measure chemical abundances for red giants in open clusters (OCs; \citealt{Reddy2012,Reddy2013,Reddy2015}). High-resolution ($R\sim$\,60,000) and high signal-to-noise (S/N) ratio optical spectra of HD 16771 were acquired during the night of 2013 November with the Robert G. Tull coud\'{e} cross-dispersed echelle spectrograph \citep{Tull1995} at the 2.7-m Harlan J. Smith reflector of the McDonald observatory. The spectra were reduced to 1D in multiple steps using various routines available in the standard spectral reduction software {\small IRAF}\footnote{IRAF is a general purpose software system for the reduction and analysis of astronomical data distributed by NOAO, which is operated by the Association of Universities for Research in Astronomy, Inc. under cooperative agreement with the National Science Foundation.}.

The wavelength range 3600 $-$ 9800 \AA\ spread across various echelle orders covered in a single 1200 s exposure is sufficient to perform an abundance analysis of many elements. The spectrum has a S/N ratio of about 220 as measured around 6000 \AA\ region and about 250 near the lithium resonance doublet at 6707.8 \AA, while at wavelengths shorter than 4000 \AA\ the S/N ratio decreases and falls to 35 around 3600 \AA\ region. We measured a heliocentric radial velocity (RV) of $+$6.9$\pm$0.1 km s$^{-1}$ which is in fair agreement with \citet{Mermilliod2008} value of $+$7.1$\pm$0.3 km s$^{-1}$. The methods of data reduction and RV measurements are described in detail in \cite{Reddy2012,Reddy2013,Reddy2015}.

Although HD 16771 as mentioned in the {\small SIMBAD}\footnote{\url{http://simbad.u-strasbg.fr/simbad/}} astronomical database is a member of the OC NGC 1039/M 34, both the proper motions ($\mu_{\alpha}cos \delta$, $\mu_{\delta}$) and RV measurements of the star differ by (+38, -11) mas/yr and 16 km s$^{-1}$ from those values measured for the cluster dwarfs \citep{JonesProsser1996}. Hence, we consider the star as a member of the Galactic disk in the cluster field.

\section{Abundance analysis}
We performed a differential abundance analysis relative to the Sun by running the {\it abfind} driver of {\scs \bf MOOG}\citep{Sneden1973phd} adopting the 1D model atmospheres \citep{CastelliKurucz2004} and the iron line equivalent widths (EWs) following our local thermodynamic equilibrium (LTE) abundance analysis technique described in \citet{Reddy2012,Reddy2013,Reddy2015}. The atmospheric parameters are derived by force-fitting the model generated iron line EWs to the observed ones by imposing the excitation and ionization equilibrium and the independence between the iron abundances and EWs. The final stellar parameters of HD 16771 derived using the iron lines are: effective temperature ($T_{\rm eff}$)=\,5050$\pm$50 K, surface gravity (log~$g$)=\,2.7$\pm$0.1 cm s$^{-2}$, microturbulence ($\xi_{t}$)=\,1.35$\pm$0.08 km s$^{-1}$ and the iron abundance [Fe/H]=\,$-0.12\pm0.05$ dex.
 
In most cases, the abundances are derived from the measured EWs but a few lines were analysed with synthetic spectra. We used a standard synthetic profile fitting procedure \citep{Reddy2015} and smoothed the computed profiles before matching to the observed ones with Gaussian profiles representing the instrumental profile, and then macroturbulence ($V_{\rm macro}$) and the rotation ($v$~sin$i$) profile of the star. The instrumental broadening was measured as the average of widths of three thorium-argon lines in the same echelle order as the stellar feature analysed by spectrum synthesis. We derived an average $v$~sin$i$ of $2.8\pm0.6$ km s$^{-1}$ and $V_{\rm macro}$\,=$3.0\pm0.4$ km s$^{-1}$ by fitting relatively isolated iron lines at 5307.4, 5141.7, 5466.9, 6187.9, 6705.1, 6739.5 and 6750.1 \AA\ following the procedure described in \citet{Carlberg2012}.  

\subsection{Li, CNO abundances and $^{12}$C/$^{13}$C ratios}
The abundances of lithium, carbon, nitrogen, oxygen as well as the carbon $^{12}$C/$^{13}$C isotopic ratios were derived by fitting synthetic spectra to selected regions of observed spectrum. As the Li resonance doublet at 6707.8 \AA\ is blended with several weak CN, Si, Ca, V and Fe features \citep{Ghezzi2009}, it is necessary to predetermine their abundances before analysing the lithium feature.

We obtained a carbon abundance of [C/H]$=\,-0.32\pm0.06$ from the analysis of C$_{2}$ Swan (1,0) lines at 5135 \AA\ and the permitted carbon line at 5380 \AA. The nitrogen abundance of [N/H]$=\,+0.25\pm0.06$ was derived by modelling of several $^{12}$C$^{14}$N lines in the 6332 \AA\ and 7938 \AA\ regions and, the oxygen abundance of [O/H]$=\,-0.08\pm0.05$ from the analysis of IR triplet (7772, 7774 and 7775 \AA) and the forbidden line at 6300 \AA. The oxygen abundance derived from the IR triplet was adjusted for the non-LTE effects by adopting empirical non-LTE corrections from \citet{Afsar2012}. Here, we referred the C, N and O abundances of HD 16771 to the solar abundances of log\,$\epsilon$(C)= 8.37 dex, log\,$\epsilon$(N)= 7.98 dex, and log\,$\epsilon$(O)= 8.83 dex derived using our linelists. 

\begin{figure}
 \begin{center}
 \includegraphics[trim=0.01cm 0.1cm 0.1cm 0.01cm, clip=true,width=0.45\textwidth,height=0.24\textheight]{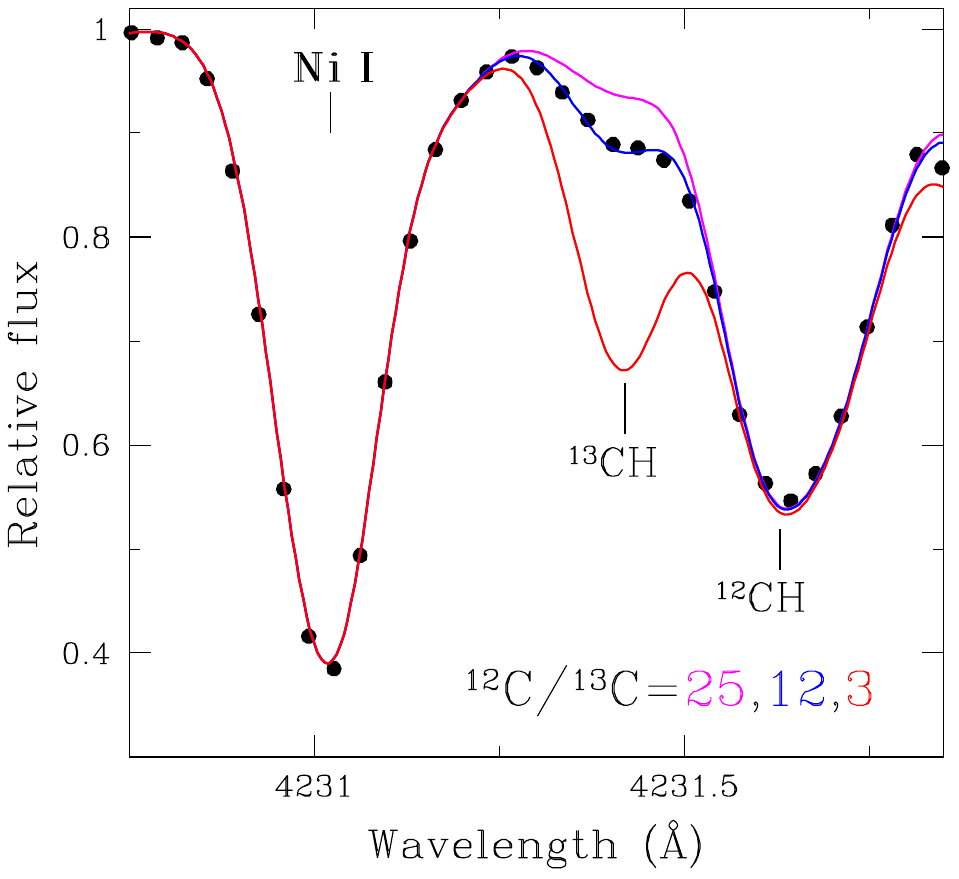}
 \caption[]{Synthetic spectra fit to the observed $^{12}$CH and $^{13}$CH features of HD 16771 around the 4231 \AA\ region for a range of $^{12}$C/$^{13}$C ratios.}
 \label{carbon_ratio}
 \end{center}
 \end{figure}

The carbon $^{12}$C/$^{13}$C isotopic ratio, which is an informative diagnostic of mixing processes \citep{Charbonnel1998} operating in the stellar atmospheres during the evolution of red giants, was measured from the synthesis of $^{12}$CH and $^{13}$CH features \citep{LamDear1972} at 4231 \AA\ (Figure \ref{carbon_ratio}). We derived a $^{12}$C/$^{13}$C ratio of 12$\pm$2, indicating that the dredge-up episodes have thoroughly mixed the surface material with the stellar interior. The molecular data for C$_{2}$ comes from \cite{Ram2014}, CH and CN data are taken from \cite{Masseron2014} and \cite{Sneden2014}, respectively.
 
\begin{figure}
\begin{center}
\includegraphics[trim=0.1cm 0.1cm 0.1cm 0.1cm, clip=true,width=0.45\textwidth,height=0.23\textheight]{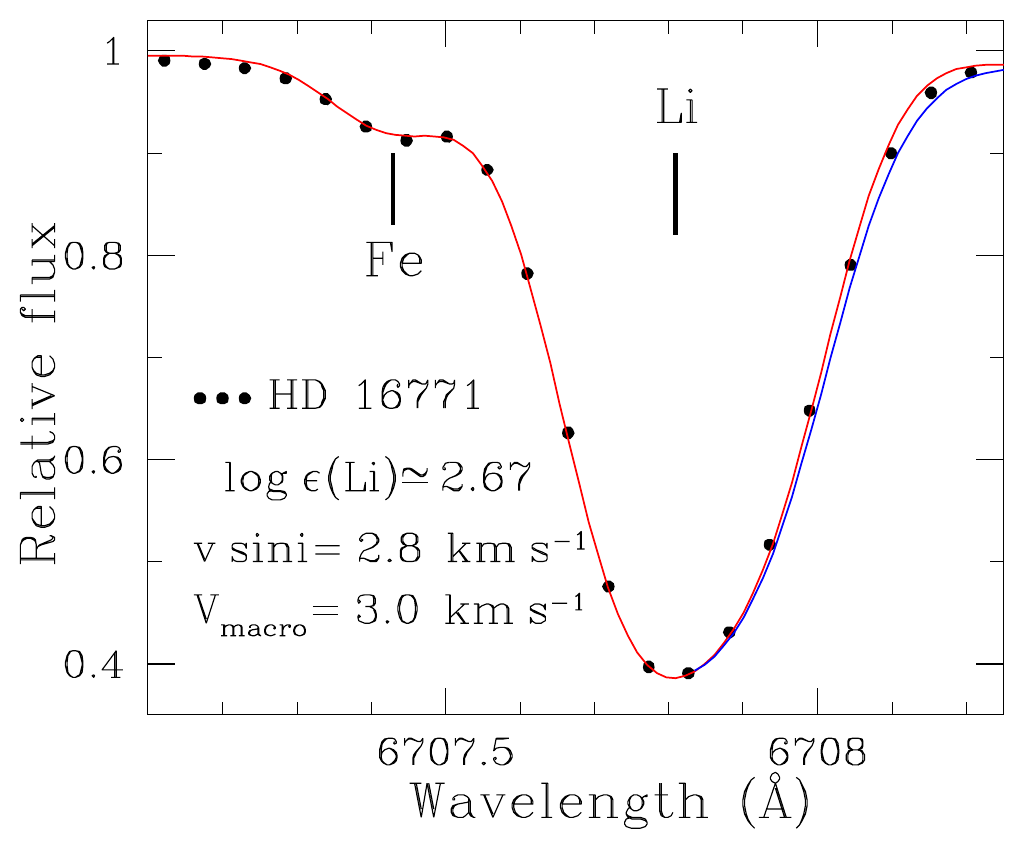} \vspace{-0.1cm}
\caption[]{The best synthetic profile fit (red) to the observed Li 6707 \AA\ line in the spectra of HD 16771. The vertical lines represent the central wavelengths of those species. Also shown is the spectrum (blue) computed for 3\% of $^{6}$Li.}
\label{lithium}
\end{center}
\end{figure}

\begin{table} 
\centering 
\caption{The chemical abundances of HD 16771 for elements from Li to Eu. The abundances measured by synthesis are presented in bold typeface while the remaining elemental abundances were calculated using the line EWs. Numbers in the parentheses indicate the number of lines used in calculating the abundance of that element. Owing to the grid boundaries, non-LTE corrections for LTE lithium abundance were applied assuming T$_{\rm eff}$=5000, log~$g$=3.0, $\xi_{t}$ =2.0 and [Fe/H]=0.0.}  \vspace{-0.2cm}
\label{abu_HD16771}
\begin{tabular}{lc}   \hline
\multicolumn{1}{c}{Species} & \multicolumn{1}{c}{HD 16771}  \\ \hline

log\,$\epsilon$(Li)$_{LTE}$  & $\bf+2.67\pm0.10(1)$  \\
log\,$\epsilon$(Li)$_{non-LTE}$ & $\bf+2.64\pm0.10$  \\
$[$C {\scs I}/Fe$]$  & $\bf-0.20\pm0.06$  \\
$[$N {\scs I}/Fe$]$  & $\bf+0.37\pm0.06$  \\
$[$O {\scs I}/Fe$]$  & $+0.04\pm0.05(5)$  \\
$[$Na {\scs I}/Fe$]$ & $+0.08\pm0.04(5)$  \\
$[$Mg {\scs I}/Fe$]$ & $+0.02\pm0.04(6)$  \\
$[$Al {\scs I}/Fe$]$ & $-0.04\pm0.02(5)$  \\
$[$Si {\scs I}/Fe$]$ & $+0.06\pm0.03(15)$  \\
$[$Ca {\scs I}/Fe$]$ & $+0.04\pm0.05(12)$  \\
$[$Sc {\scs II}/Fe$]$& $\bf+0.10\pm0.04(2)$  \\
$[$Ti {\scs I}/Fe$]$ & $+0.01\pm0.05(16)$  \\
$[$Ti {\scs II}/Fe$]$& $-0.05\pm0.04(7)$  \\
$[$V {\scs II}/Fe$]$ & $+0.05\pm0.07(14)$  \\
$[$Cr {\scs I}/Fe$]$ & $+0.02\pm0.04(10)$  \\
$[$Cr {\scs II}/Fe$]$& $+0.04\pm0.05(6)$  \\
$[$Mn {\scs I}/Fe$]$ & $\bf-0.08\pm0.07(2)$  \\
$[$Fe {\scs I}/H$]$  & $-0.12\pm0.05(163)$  \\
$[$Fe {\scs II}/H$]$ & $-0.13\pm0.05(15)$  \\
$[$Co {\scs II}/Fe$]$& $+0.05\pm0.04(6)$  \\
$[$Ni {\scs I}/Fe$]$ & $-0.08\pm0.04(27)$  \\
$[$Cu {\scs II}/Fe$]$& $\bf-0.13\pm0.04(1)$  \\
$[$Zn {\scs I}/Fe$]$ & $\bf-0.03\pm0.07(1)$  \\
$[$Y {\scs II}/Fe$]$ & $+0.05\pm0.05(5)$  \\
$[$Zr {\scs I}/Fe$]$ & $+0.17\pm0.06(4)$  \\
$[$Ba {\scs II}/Fe$]$& $\bf+0.27\pm0.08(1)$  \\
$[$La {\scs II}/Fe$]$& $+0.08\pm0.04(4)$  \\
$[$Ce {\scs II}/Fe$]$& $+0.12\pm0.04(4)$  \\
$[$Nd {\scs II}/Fe$]$& $+0.23\pm0.05(13)$  \\
$[$sm {\scs II}/Fe$]$& $+0.23\pm0.05(7)$  \\
$[$Eu {\scs II}/Fe$]$& $\bf+0.17\pm0.05(1)$  \\

\hline
\end{tabular} \vspace{-0.2cm}
\end{table}

\begin{table} 
\centering 
\caption{Summary of the stellar parameters of HD 16771.}  \vspace{-0.2cm}
\label{stellar_param}
\begin{tabular}{lcc}   \hline
\multicolumn{1}{c}{Parameter} & \multicolumn{1}{c}{value} & \multicolumn{1}{c}{Reference}  \\ \hline
 V (mag)     &  7.34   &  Nicolet (1978)  \\
B-V (mag)    &  0.94   &  Nicolet (1978)  \\
parallax (mas)  &  $3.95\pm0.15$ &  This work \\
distance (pc)   &  $253\pm10$    &  This work       \\ 
$RV_{helio}$ (km s$^{-1}$) & $+6.91\pm0.13$  &  This work   \\
T$_{\rm eff}$ (K)          &  $5050\pm50$    &  This work  \\
log~$g$ (cm s$^{-2}$)      &  $2.7\pm0.1$    &  This work  \\
$\xi_{t}$ (km s$^{-1}$)    & $1.35\pm0.08$   &  This work  \\
$[$Fe/H$]$ (dex)           & $-0.12\pm0.05$  &  This work  \\
$v$~sin$i$ (km s$^{-1}$)   & $2.8\pm0.6$     &  This work  \\
\hline
Age (Gyr)                  & $0.76\pm0.13$  &  This work  \\
$M/M_{\odot}$      & $2.4\pm0.1$  &  This work  \\
$R/R_{\odot}$      & $11.1\pm0.5$   &  This work  \\
log\,($L/L_{\odot}$)& $1.92\pm0.09$  &  This work  \\
 
\hline
\end{tabular}
\end{table}

Thereafter, we measured the lithium abundance from the synthesis of resonance doublet at 6707.8 \AA\ adopting the hyperfine structure data from \cite{Ghezzi2009}. A fit to the observed lithium profile without the need for $^{6}$Li, as shown in Figure \ref{lithium}, is obtained for a LTE abundance of log\,$\epsilon$(Li)$=+2.67\pm0.10$ dex. Our LTE lithium abundance was corrected for the non-LTE corrections provided by \cite{Lind2009} for a star of given $T_{\rm eff}$, log~$g$, $\xi_{t}$ and [Fe/H].

As HD 16771 has no measured stellar parallax ($\pi$), initial guess of $\pi$ was made by incorporating the derived spectroscopic atmospheric parameters into the known relation 
\noindent \vskip1ex
$\log\pi$= 0.5\,[$\log(g$/${g}_{\odot})-$$\log(M/M_{\odot})-$4\,$\log(T_{\rm eff}/T_{\rm eff}{\odot})$]
\begin{equation}
 -0.2(V-A_{V}+BC_{V}+0.25)  
\end{equation}
where A$_{V}$ is the interstellar extinction in V band and the bolometric correction $BC_{V}$ is estimated from the spectroscopic T$_{\rm eff}$ and [Fe/H] using Alonso et al's (1999) calibration. Here we adopt log~$g_{\odot}$= 4.44 cm s$^{-2}$ and T$_{\rm eff},_{\odot}$= 5777 K and assume that HD 16771 has a mass of M$=$ 2.0 $M_{\odot}$. 

Using the Johnson V magnitude along with the spectroscopic T$_{\rm eff}$ and [Fe/H], our guess of $\pi=$ 4.6 mas is varied until the surface gravity estimate from the PARAM\footnote{\url{http://stev.oapd.inaf.it/cgi-bin/param_1.3}} code \citep{daSilva2006} and PARSEC stellar evolutionary tracks \citep{Bressan2012} matches our spectroscopically determined value of log~$g$, i.e., 2.7 cm s$^{-2}$. We adopted the stellar mass and age of HD 16771 from the isochrone that corresponds to our best measured value of $\pi=3.95\pm0.15$ mas. 

The spatial velocities (U$_{LSR}$, V$_{LSR}$, W$_{LSR}$)$=$(-24.9, -29.8, 10.0) km s$^{-1}$, distance away from the Galactic plane z$=$0.07 kpc, and a thin disk membership probability (Reddy et al. 2015) of 98\% along with the chemical composition (see Table \ref{abu_HD16771}) of HD 16771 agree well for an object typical from the Galactic thin disk. A summary of stellar parameters of HD~16771 is presented in Table \ref{stellar_param}.

\section{Discussion}

\begin{figure}
\begin{center}
\includegraphics[trim=0.1cm 0.2cm 0.1cm 0.01cm, clip=true,width=0.45\textwidth,height=0.30\textheight]{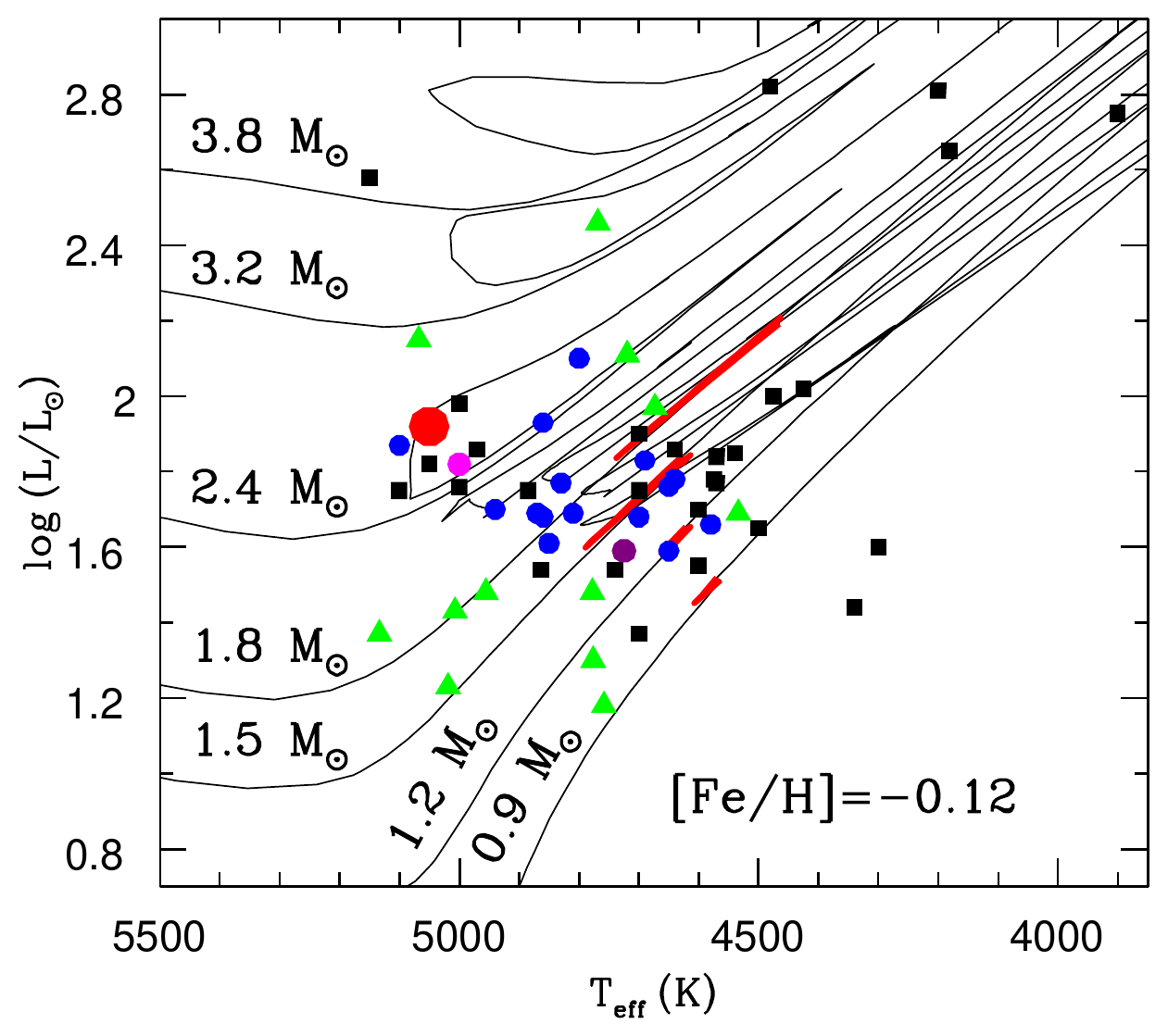}
\caption[]{Position of HD 16771 (red) in the log(L/L$_{\odot}$) versus T$_{\rm eff}$ plane along with other Li-rich giants (see text). The location of the LFB is indicated by red stripes. HD~16771 with T$_{\rm eff}$=5050 K is clearly on the high temperature side of the bump found for low-mass stars where the possibility of extra-mixing is proposed \citep{CharBal2000}. }
\label{cmdHD16771}
\end{center} \vskip-0.2cm
\end{figure}

According to the PARSEC evolutionary tracks \citep{Bressan2012} and a Bayesian estimation method \citep{daSilva2006}, HD 16771 at a distance of $253\pm10$ pc from the Sun (see Table \ref{stellar_param}), and with a mass M\,=\,$2.4\pm0.1\,M_{\odot}$ and age\,=\,$0.76\pm0.13$ Gyr shares the region enjoyed by the He-core burning clump giants that have evolved to the clump via the RGB tip in the HR$-$diagram. The position of HD 16771 (large red dot) and the other Li-rich giants extracted from {\citet[green]{Adamow2014,Adamow2015}, \citet[magenta]{SilvaAguirre2014}, \citet[purple]{Jofre2015}, and from \citet[blue]{Kumar2011} and additional stars collected by Kumar et al. from the literature (black squares) are shown in the HR$-$diagram (Figure \ref{cmdHD16771}) along with the PARSEC evolutionary tracks computed for stars in the mass range from 0.9 to 3.8 M$_{\odot}$ from \cite{Bressan2012}.

The lithium content of log\,$\epsilon$(Li$_{non-LTE}$)$=+$2.64 dex in the atmosphere of HD 16771 declares the star to be a Li-rich giant but can not be readily explained with Li synthesis occurring during the LFB or at the He-core flash. HD 16771 with a mass M\,=\,$2.4\pm0.1\,M_{\odot}$ has no chance of encountering the LFB as the RGB finishes \citep{SacBoo1999} before the H-burning shell reaches the $\mu$-barrier and thus no extra-mixing process is expected to trigger Li production via cool bottom process \citep{SacBoo1999}. Also it should not have experienced He-core flash at the RGB tip as the star with mass M\,$\geq$\,2.2\,M$_{\odot}$ starts burning He in its core \citep{CharLag2010} while on the RGB. Moreover, the star can not have operated the hot bottom burning mechanism \citep{SacBoo1992} probably responsible for Li production in stars high on the Asymptotic Giant Branch (AGB). Even allowing for errors in the ingredients used in the estimation of stellar mass, the location of HD~16771 on the HR diagram supports these conclusions. 

In this vein, we note the detection of warm Li-rich clump-giants with masses above 2.2\,M$_{\odot}$ in two different studies: \citet{MartellShetrone2013} have identified a sample of clump-giants with the combination of T$_{\rm eff}$=5000-5200, log~$g$=2.6-3.1 and $^{12}$C/$^{13}$C$\leq$15. The study of \citet{Brown1989} has included a clump-giant HD~120602 with T$_{\rm eff}$=5000, log~$g$=3.00 whose $^{12}$C/$^{13}$C has been measured at 16 \citep{BerdSava1994}. HD~16771 with $^{12}$C/$^{13}$C=12$\pm$2 is another such example whose Li enrichment is not linked to the episode of extra-mixing near the LFB or He-core flash as suggested previously for stars of mass M$<$2.2\,M$_{\odot}$ \citep{CharBal2000,Kumar2011}. Here, we avoid plotting the sample of \citet{MartellShetrone2013} in Figure \ref{cmdHD16771}, as they have no measured parallaxes or V magnitudes.

However, the observed low values of C/N$\sim$0.7 and $^{12}$C/$^{13}$C$\sim$12, although far from the predictions of conventional mixing (\citealt{Charbonnel1998}), clearly show that HD 16771 has experienced non-canonical mixing and thus the possibility of preserving initial main-sequence lithium content and external pollution from a companion is excluded. Moreover, the external pollution from an evolved Li-rcih AGB companion is associated with the enhancements in the $s$-process elements, characteristics not carried by HD~16771. Together five epochs of RV observations collected from \citet{Mermilliod2008} show no sign of binarity.  

Thus, the lithium overabundance in HD 16771 requires an alternative to models of extra-mixing (see \citealt{Carlberg2012} and references therein) that invoke destruction of $\mu$-barrier in solar metallicity stars with degenerate He cores (M$<$\,2.2 M$_{\odot}$) that have evolved through both the LFB and He-core flash stages. However, for high mass stars (M$>$\,2.2\,M$_{\odot}$) with non-degenerate He cores, Li evolution ceases once the star completes first dredge-up. 

If fresh $^{7}$Be is synthesized via Cameron-Fowler mechanism in the interior of a star like HD 16771, the mixing should be fast enough to carry $^{7}$Be (and $^{7}$Li) to cool layers where it decays to $^{7}$Li. Indeed, \citet{Denissenkov2012} proposed a stirring mechanism in the RGB phase of low-mass stars with fast internal rotation which leads to a temperature gradient between the hydrogen-burning shell and the envelope that permits enhanced mixing across the radiative zone. As a consequence, a low-mass star make extended zigzags toward the position of the red-clump star, before resuming its ascent along the RGB. As the theoretical models do not predict Li production during the He-core flash, \citet{Denissenkov2012} proposed that the red-clump stars in the study of \citet{Kumar2011} are LFB low-mass stars which had made their excursion in the HR-diagram towards luminosities compatible with the expected location of red-clump stars. However, our star has a very low probability to encounter such scenario and sits on the high temperature side and far away from the LFB found for low mass stars (Figure \ref{cmdHD16771}).

Several other sources of enhanced extra-mixing have been suggested to explain the presence of Li-rich giants all along the RGB. Among those are \citet{DenissenkovWeiss2000} and \citet{DenissenkovHerwig2004} models of activation of $^{7}$Be-transport mechanism followed by the engulfing of 
a giant planet \citep{Alexander1967}. The engulfing of close-in planets might cause shear instabilities at the base of the convective envelope that could turn-on the dynamo action \citep{SiessLivio1999}. The associated thermohaline \citep{CharZahn2007} and magneto-thermohaline mixing \citep{Denissenkov2009} could circulate the material fast enough to replenish Li in the stellar envelope that also lowers the $^{12}$C/$^{13}$C ratio from its first dredge-up value.

Exoplanets observations show deficiency of close-in planets around evolved stars compared to their main sequence counterparts, for which planet engulfment is likely a plausible explanation \citep{Villaver2014}. In principle, HD 16771 has evolved enough to engulf a putative close-in planet, 
as in known hot Jupiter systems, for which the semi major axis is usually less than 0.05 AU \citep{Adamow2015}. On the account of positive detection of Li excess in HD 16771 at the red-clump phase, we speculate that the engulfing of planet would have occurred during RGB stage, where the giant's outer envelope expands and reaches to its maximum stellar radius at the RGB tip. And for stars with $M=2.4\,M_{\odot}$ the critical initial semi-major axis for engulfment is $\sim$\,0.12-0.18 AU and for giant stars in the age range 620$-$650 Myr \citep{Delgado2016}. Therefore, HD 16771 with an age of 760$\pm$130 Myr would have undergone extra-mixing triggered by the engulfment of planet(s) probably 100 Myr ago and yet the observed Li overabundance suggest that the elapsed time is insufficient for Li dilution. Also the rapid evolution from the RGB to red-clump phase and the shallow convective envelope of the He-core burning stars could aid in preserving Li as the envelope material hardly circulates to the hot internal layers where Li is promptly destroyed \citep{Delgado2016}. 

\citet{SiessLivio1999} have suggested that engulfment of substellar objects not only raise the star's $^{7}$Li abundance but could also trigger a prompt mass-loss, spin-up of the star, and the possible generation of magnetic fields, depending on the masses of both star and planet and on the planet’s orbital parameters. The slow rotation of 2.8 km s$^{-1}$, typical for an evolved giant, suggest that HD 16771 could have undergone efficient rotational braking in this time frame and carried away most of the envelope's angular momentum in a mass-loss event. Others who have considered planet accretion as the solution to Li enhancement in some giant stars include \citet{Adamow2012}, \citet{Carlberg2012} and \citet{Delgado2016}. A major fraction of Li-rich giants with planets in those studies present $v$~sin$i\,<$ 4 km s$^{-1}$, and our giant would be one such candidate showing a slow rotational velocity. These results, as previously noted by \citet{FekelBal1993}, weaken the claim that planet ingestion, high lithium abundance and rapid rotation are related inextricably, as the timescales for Li enrichment are not necessarily the same as the timescales for rotational spin-up and spin-down, or for increased or decreased Ca~II H and K emission. 

A thorough investigation of IR fluxes extracted from 2MASS \citep{Cutri2003} and WISE \citep{WISE2010} reveals no sign of IR excess at short wavelengths (i.e., J, K, and WISE[3.4, 4.6, 12 and 22 $\mu$m]). Using the test given by de La Reza et al. (1996), which makes use of the [60-25] and [25-12] IRAS data, we confirm that HD 16771, in common with BD+48~740 \citep{Adamow2012}, shows IR excess at 60 $\mu$m, suggesting that the engulfment happened some time ago allowing the shell time to expand and cools down. Supporting this assertion are the lack of observations of asymmetric H${\alpha}$ and irregular Na D lines in our spectra, which are expected as a result of very recent event of intense mass-loss. Computations show that the dust shell expansion to produce excess emission at 60 $\mu$m will take about 10$^{4}$ yr \citep{delaReza1996}, which is consistent with the proposal that mass-loss would have carried away most of the envelope's angular momentum and, eventually, slow down HD 16771 to the observed value of $v$~sin$i=$2.8 km s$^{-1}$ in the red-clump phase. Along these lines we speculate that HD 16771 is in the cool extended dust shell and high lithium abundance phase of Siess \& Livio's (1999) scenario.

Another signature of planet engulfment is the detection of $^{9}$Be enrichment in stars. A search for beryllium enhancements in a limited sample of Li-rich and Li-normal giants with T$_{\rm eff}$=\,3900$-$4800 K and masses in the range 1.4-3.2\,M$_{\odot}$ provides no detection of the $^{9}$Be line at 3131 \AA\ \citep{Melo2005}. Although, the Be~{\scs II} line at 3131 \AA\ falls outside our spectral range to measure its abundance, its usefulness for deriving Be abundances decreases with temperature below 5400 K to a point where it will be useless \citep{Santos2004}. The source of difficulty arises from the severe blending of Be line with other atomic lines and their dominance at low temperatures around 3131 \AA, and the missing near-UV opacity which is not needed at the solar effective temperature (see \citealt{Santos2004} and references therein). Therefore, the Be abundance determinations requires further scrutiny.

We note here that engulfing a putative close-in planet(s) with an equivalent mass of 30 Jupiters (using equation 2 in \citealt{SiessLivio1999}), whose atmosphere was hydrogen-poor (say by about 1.5 dex) due to the evaporation of hydrogen atoms by the sustainable radiation pressure from the host star \citep{Alexander1967,VidalMadjar2003}, into a reasonable convective envelope of 80\% the mass of the red giant \citep{Brown1989} would raise the lithium abundance of HD~16771 to 2.8 dex and the beryllium abundance by 1.0 dex without affecting appreciably the abundances of other elements. The above calculations are made for a giant star whose convective envelope is void of both the lithium and beryllium abundances. However, the true enhancement in the beryllium abundance would be much smaller (about 0.4 dex if we assume a beryllium abundance of 0.8 dex on the RGB) and probably undetectable because of severe blending of Be line with other atomic lines and the difficulty in continuum placement in those crowded regions of the spectra. Then a prompt depletion of $^{6}$Li and a progressive dilution of $^{7}$Li would naturally explain Li overabundance and IR excess in HD~16771 without any need for Cameron-Fowler mechanism. 
Therefore, the absence/presence of $^{9}$Be overabundance in HD~16771 would not necessarily contradict the engulfment scenario, as the close-in planets are fated to undergo regular radiational harassment from their host stars which would eventually evaporate hydrogen much faster than other elements (\citealt{VidalMadjar2003,Bourrier2014}). 

The puzzling overabundance of lithium in HD 16771 makes it an oustanding prospect for future studies to set strong constraints on the role of planet engulfment and the episode of Li synthesis for stars that neither experience LFB nor the He-core flash, and on the timescales for Li dilution, rotational spin-up and spin-down of stars.
 
\vskip1ex 
{\bf Acknowledgements:}
 
We are grateful to the McDonald Observatory's Time Allocation Committee for granting us observing time for this project. DLL wishes to thank the Robert A. Welch Foundation of Houston, Texas for support through grant F-634. We are grateful to the anonymous referee for a very careful and constructive report that led to improvements of the Paper.

This research has made use of the WEBDA database, operated at the Department of Theoretical Physics and Astrophysics of the Masaryk University and the NASA ADS, USA. This research has made use of the SIMBAD database and Aladin sky atlas, operated at CDS, Strasbourg, France. This publication makes use of data products from the Two Micron All Sky Survey, which is a joint project of the University of Massachusetts and the Infrared Processing and Analysis Center/California Institute of Technology, funded by the National Aeronautics and Space Administration (NASA) and the National Science Foundation (NSF).

\bibliographystyle{aa} 
\bibliography{ms.bib}  

\end{document}